# Mauna Kea Spectroscopic Explorer (MSE): a preliminary design of multi-object high resolution spectrograph

Kai Zhang[*a,b], Yifei Zhou[a,b,c], Zhen Tang[a,b], Will Saunders[d], Kim A. Venn[e], Jianrong Shi[f], Alan W. McConnachie[g,h], Kei Szeto[g], Lei Wang[a,b], Yongtian Zhu[a,b], Zhongwen Hu[a,b]

a. National Astronomical Observatories / Nanjing Institute of Astronomical Optics & Technology, Chinese Academy of Sciences, Nanjing 210042, China
b. Key Laboratory of Astronomical Optics & Technology, Nanjing Institute of Astronomical Optics & Technology, Chinese Academy of Sciences, Nanjing 210042, China
c. University of Chinese Academy of Sciences, Beijing 100049, China
d. Australian Astronomical Observatory, North Ryde, NSW 2113, Australia
e. University of Victoria, Victoria, BC V8P 5C2, Canada
f. National Astronomical Observatory of China, Chinese Academy of Sciences, Beijing 100012, China
g. CFHT Corporation, Kamuela, Hawaii 96743, USA
h. National Research Council of Canada, Victoria, BC, V9E 2E7, Canada

**ABSTRACT**

The Maunakea Spectroscopic Explorer (MSE) project will transform the CFHT 3.6m optical telescope to a 10m class dedicated multi-object spectroscopic facility, with an ability to measure thousands of objects with three spectral resolution modes respectively low resolution of R≈3,000, moderate resolution of R≈6,000 and high resolution of R≈40,000. Two identical multi-object high resolution spectrographs are expected to simultaneously produce 1084 spectra with high resolution of 40,000 at Blue (401-416nm) and Green (472-489nm) channels, and 20,000 at Red (626-674nm) channel. At the Conceptual Design Phase (CoDP), different optical schemes were proposed to meet the challenging requirements, especially a unique design with a novel transmission image slicer array, and another conventional design with oversize Volume Phase Holographic (VPH) gratings. It became clear during the CoDP that both designs presented problems of complexity or feasibility of manufacture, especially high line density disperser (general name for all kinds of grating, grism, prism). At the present, a new design scheme is proposed for investigating the optimal way to reduce technical risk and get more reliable estimation of cost and timescale. It contains new dispersers, F/2 fast collimator and so on. Therein, the disperser takes advantage of a special grism and a prism to reduce line density on grating surface, keep wide opening angle of optical path, and get the similar spectrum layout in all three spectral channels. For the fast collimator, it carefully compares on-axis and off-axis designs in throughput, interface to fiber assembly and technical risks. The current progress is more competitive and credible than the previous design, but it also indicates more challenging work will be done to improve its accessibility in engineering.

______________________________________________

* kzhang@niaot.ac.cn, phone: +86-25-85482316

# 1. INTRODUCTION

The Maunakea Spectroscopic Explorer (MSE) project will transform the CFHT 3.6m optical telescope to a 10m class dedicated multi-object spectroscopic facility, with an ability to simultaneously measure thousands of objects with three spectral resolution modes respectively low resolution of R≈3,000, moderate resolution of R≈6,000 and high resolution of R≈40,000 [1] [2]. Two identical multi-object high resolution spectrographs (HR) are expected to simultaneously produce 1084 spectrum with high resolution of 40,000 at Blue (401-416nm) and Green (472-489nm) channels, and 20,000 at Red (626-674nm) channel. There is a requirement to be able to observe at the resolution of 40,000 in any working window with narrow bandpass of 1/30 over the wavelength range of 360-500nm, and observe at the resolution of 20,000 in any working window with bandpass of 1/15 over the wavelength range of 500-900nm.

This capability enables detailed study of weak spectral lines in crowded regions, which is essential to probe various chemical species that provide key insights into the evolution of the Galaxy and the formation of the elements [3]. The ESA space satellite Gaia is dedicated to probing the properties of all stars brighter than G=20 magnitudes, and here MSE will be the ultimate Gaia follow-up facility, in particular, which decomposes the outer regions of the Galaxy into its constituent star formation events by accessing a range of chemical tracers that sample a large number of nucleosynthetic pathways. PLATO is another ESA mission developed in order to monitor a large number of bright stars to search for planets. MSE will provide spectroscopic characterization at high resolution and high signal to noise ratio (SNR) of the faint end of the PLATO target distribution (g~16), to allow for statistical analysis of the properties of planet-hosting stars as a function of stellar and chemical parameters.

The MSE project completed all the conceptual design work early 2018 [4], including the successful Conceptual Design Review (CoDR) for the high resolution spectrograph (HR) in April of 2017. The Sensitivity Budget Allocation Document clearly declares the sensitivity requirement at high resolution for the complete observing system from atmosphere to the detector. The MSE-HR shall have a SNR per resolution element at a given wavelength that is greater than or equal to 10 for a 1-hour observation of a point source with a flux density of 3.6e-28 ergs/sec/cm$^2$/Hz at that wavelength, for all wavelengths in the relevant window longer than 400nm. Between 370-400nm, the SNR shall not be less than 5 at any wavelength in the relevant window. The observing condition in which this requirement shall be met correspond to a sky brightness of 19.5mags/sq.arcsec in the V-band at an airmass of 1.2, and a delivered image quality at that airmass of 0.6 arcsec Full Width at Half Maximum (FWHM) in the g band. Converted from the requirement in spectral resolution and sensitivity, the system throughput requirement is double over the wavelength range of 400-500nm, see the cyan asterisk curve in Figure 1, and the spectrograph throughput is required being no lower than the solid curves (Blue/Green/Red), which is derived by excluding the other contributions (atmosphere, telescope, Fiber Transmission System (FiTS)) [5].

At the Conceptual Design Phase (CoDR), some optical design schemes were investigated to meet the science requirement, especially a unique design with a novel transmission image slicer array [6], and another conventional design with oversize Volume Phase Holographic (VPH) gratings. The former design is capable of reducing pressure on optical aperture and grating efficiency at the cost of doubling the instrument number and adding a complicated pre-optics. The latter design adopted a conventional scheme for Multi-Object Spectrograph (MOS) to give up the using of image slicer, and try to use a mosaic grating instead of the ultra-high line density grism (ld>6,000 l/mm). Refer to the HERMES installed on AAT [7], it suffers from serious polarization phenomena on diffraction and AR coatings generated by big incident angle ($α ≈ 68$ degree). These two designs highlight the importance of disperser (general name for all kinds of grating, grism, prism), including diffraction efficiency, mosaic grating technology and manufacture feasibility. In Figure 1, thick solid lines (Blue/Green/Red) represent the disperser requirements derived by the estimated efficiency of spectrograph optics at the CoDP. The diffraction efficiency shall be at least higher than 70% at the wavelength shorter than 500nm. It's very challenging for the ultra-high line density disperser (ld≥6,000 l/mm) according to the technical level in fabrication.

In 2018, we actively studied in trade-off study on a new design, Light Preliminary Design Phase (LPDP), in order to reduce technical risk and obtain more reliable estimation in cost and timescale. The current design is elaborated at the following sections, respectively section 2 describes the scheme to achieve the throughput requirement, section 3 explains the current disperser design, section 4 describes the optical design, excluding the disperser, and finally section 5 concludes the knowledge obtained in the current study and describes some ideas for the ongoing work.

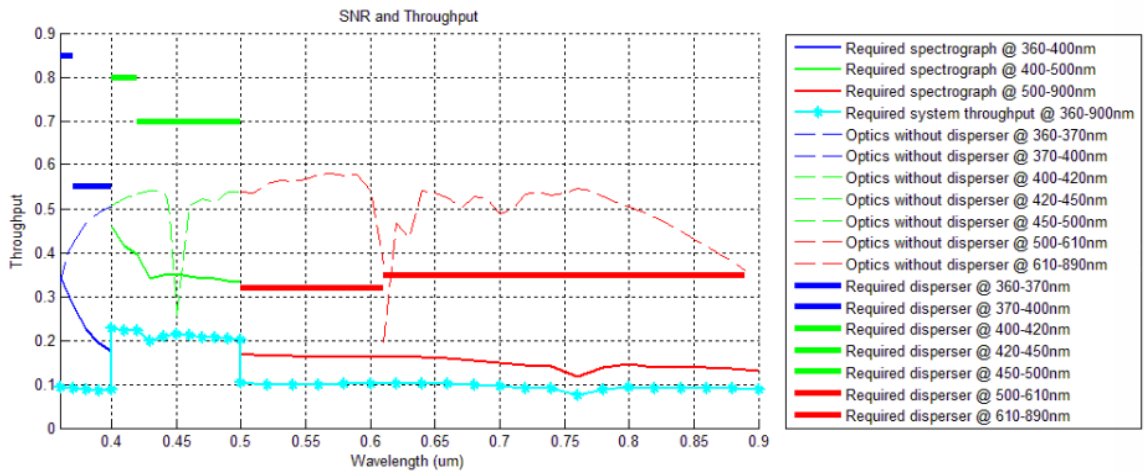

Figure 1. Throughput requirement of the MSE-HR

## 2. DESIGN SCHEME

From the above introduction, it's clear that the instrument functionality and its sensitivity are the most important criteria to make the optimal optical design, and ultra-high line density disperser is the most risky optical part affecting the spectrograph throughput. The relevant technical requirement is generalized as the below:

(1) Two identical HR spectrographs accommodate 1084 fibers in total, namely 542 fibers per spectrograph.

(2) Three spectral working windows (BWW-01/GWW-01/RWW-01) are required for the first round of spectral survey, respectively 408.55nm with bandpass of 1/30 (BWW-01), 481nm with bandpass of 1/30 (GWW-01), 650.5nm with bandpass of 1/15 (RWW-01). And it shall provide the feasibility to change working windows over the whole wavelength range of 360-900nm.

(3) Based on the nominal fiber diameter of Φ0.8arcsec, resolution shall be in range of 38,000 – 42,000 at the wavelength shorter than 500nm, and no point is lower than 35,000; Resolution shall be in range of 18,000 – 22,000 at the wavelength longer than 500nm.

(4) According to the sensitivity requirement and conceptual estimation of throughput, the overall throughput of spectrograph shall be no lower than 30% in the wavelength range of 360 – 400nm, no lower than 40% in the wavelength range of 400 – 500nm, no lower than 18% in the wavelength range of 500 – 900nm.

It's critical for the design at the LPDP to increase the disperser efficiency and ease manufacture by lowering line density while maintaining acceptable spectral resolution. The instrument resolution depends on serval factors, including telescope effective aperture, fiber diameter, the optical system and disperser. Besides of improving the disperser itself, it's more important to achieve the goal by sharing performance pressure among the other factors.

Fiber diameter is a direct factor to affect the instrument resolving power. It was reduced from $\phi$1arcsec to $\phi$0.8arcsec after further science analysis durung the CoDP. At the present, it is considered if smaller fiber diameter could be used to relieve the dispersion pressure by 5%-10% at the cost of limited SNR loss. Simulation in Figure 2 (a) shows that $\phi$0.75arcsec fiber possibly yields averaged SNR loss of 5% at the different telescope zenith angles (0 degree, 30 degree, 50 degree). It looks more reasonable than 10% loss led by $\phi$0.70arcsec fiber. So the current design chooses a $\phi$0.75arcsec fiber to relieve dispersion pressure by 6.25%, and prepare a backup scheme to compensate for its SNR loss. The backup scheme replaces the $\phi$0.75arcsec fiber by the combination with $\phi$1.0arcsec fiber and a 0.75arcsec slit [7]. It results in a fiber area of 30% bigger than $\phi$0.75arcsec fiber, see the shaded zone in Figure 2 (b), to compensated the loss in stellar light acquisition due to smaller fiber diameter.

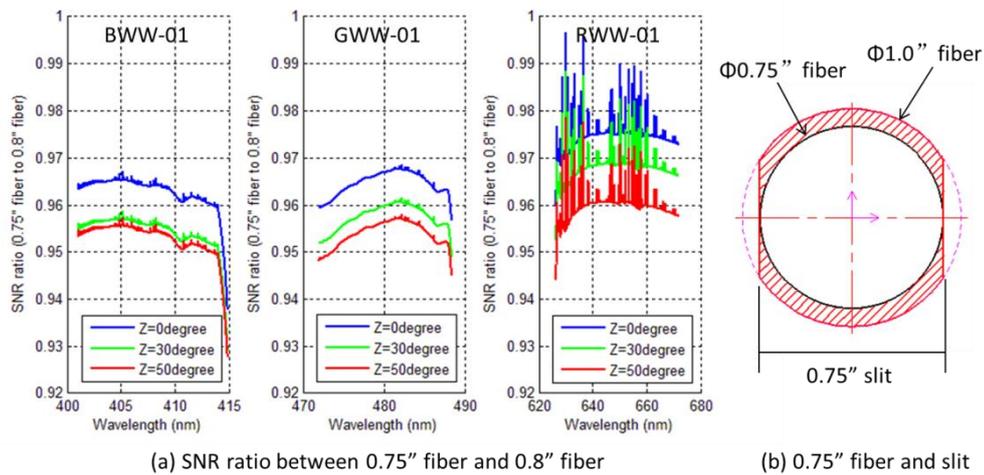

(a) SNR ratio between 0.75" fiber and 0.8" fiber      (b) 0.75" fiber and slit

Figure 2. Fiber selection

At the telescope prime focus, the nominal focal ratio is F/1.926 corresponding to the full telescope aperture of $\phi$11.25m. At the spectrograph entrance, it's desirable to increase the throughput and simplify the optics by avoiding any pre-optics for de-magnification. Among some existing MOSs, their focal ratios of collimator without pre-optics are distributed around F/3, see Table 1. It's very challenging to design a much faster collimator, especially with multiple arms, and preferably with unobstructed pupil. A collimator operating at f/1.926 would restrict losses due to Focal Ratio Degradation (FRD) to 5%, see Figure 3 [8]. According to the test result shown in Figure 3, it's reasonable for the HR collimator to use a slightly slower focal ratio around F/2 at the limited cost of throughput (7% - 10%, including FRD=5%), which is lower than the throughput loss generated by F number slowing pre-optics. Benefited from slower F number, the corresponding effective telescope aperture becomes smaller. It indirectly results in relieving the dispersion pressure by ~5%.

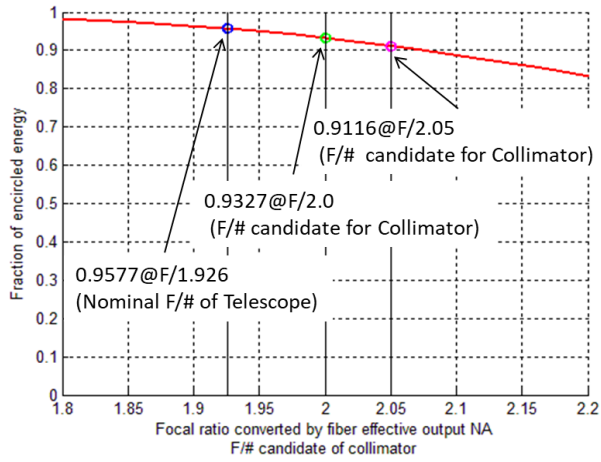

Figure 3. Test of fiber throughput [8]

Table 1. Capability of some existing MOSs

| Spectrograph | PFS [9] | MOONS [10] | WEAVE [11] | 4MOST [12] | HERMES [7] |
| --- | --- | --- | --- | --- | --- |
| Telescope | Subaru [8.2m] | VLT [8.2m] | WHT [4.2m] | VISTA [4m] | AAT [3.9m] |
| Focus [F/#] | Prime [F/2.8] | Nasmyth [F/3.5] | Prime [F/3.1] | Cassagrain [F/3] | Prime [F/3.15] |
| Waveband | 0.38-1.3um | 0.65-1.8um | 0.37-0.96um | 0.36-0.95um | 0.37-1.0um |
| LM mode | 2,000-5,000 | 4,000-6,000 | 5,000 | >4,000/>6,000 | — |
| HR mode | — | 9,200(I)/19,000(J/H) | 20,000 | >18,500 | 28,000/>40,000 |
| Fiber diameter | 1.1" | 1.05" | 1.3" | 1.45" | 2.35" |
| Number of fibers | 600/spec. | 512/spec. | 964/940/spec. | 812+10/spec. | 392/spec. (x2) |
| Number of spec. | 4 | 2 | 1 | 2 | 1 |
| Collimated aperture | Φ 280mm | Φ 265mm | Φ 190mm | Φ 200mm | Φ 190mm |
| Collimator | F/2.5 on-axis | F/3.5 on-axis | F/3.1 off-axis | F/3 off-axis | F/6.32 off-axis |
| Disperser | VPH grating | VPH grism+Prism | VPH grating | VPH grating+Prism | VPH grating |
| Camera | F/1.1 reflective | F/1.04 reflective | F/1.8 transmission | F/1.77 transmission | F/1.7 transmission |
| Detector | 4Kx4K@15um | 4Kx4K@15um | 6Kx6K@15um | 6Kx6K@15um | 4Kx4K@15um |
| Image size | 3.8pixels | 3pixels | 3.2pixels | 3.3pixels | 5pixels |

At the design aspect of the disperser, the grism is preferred because it provides more variables than grating to optimize the system parameters, and effectively avoid the polarization phenomena on AR coating. The instrument resolving power is reflected by total line number (N) on grating surface, in essence, w the product of line density and grating area, see Equation 1. Normally, grating area depends on collimated aperture (Dc) sand incident angle on grating surface ($α$). Based on this, the current design follows the same collimated aperture of spectrograph as the conceptual design, Φ

300mm. So it requires total line number (N) in the medium of Fused Silica shall be no less than 3.2million lines in the BWW-01, 2.7million lines in the GWW-01, and 0.95million lines in RWW-01. In addition, it shall be paid attention that a special disperser design introduced at Section 3 enables to increase grating area by ~10% in dispersed dimension (Y) to reduce the line density.

$$N = (R \times \Phi \times D_t)/(r_{fn} \times \lambda) \tag{1}$$

Where, R is spectral resolution, Φ is fiber diameter or slit width, in unit of arcsec, $D_t$ is nominal telescope aperture of Φ11.25m, $r_{fn}$ is the gain of focal ratio obtained from the collimator, λ is the central wavelength in a single working window.

Therefore, it's feasible to reduce the line density by ~7%, because it is not linear relationship among these variables, see Table 2. Although the ability to increase in disperser efficiency is limited, the design scheme increases the spectrograph throughput by optimizing the optical design of the catadioptric collimator and transmission camera. Compared with the conceptual design, see Table 3, the current design makes great gain to increase the spectrograph throughput, and get better instrument layout for the mechanics (Section 4). Compared with the existing MOSs shown in Table 1, it's obvious that F/2.05 off-axis collimator and F/1.55 transmission camera will be more challenging for glass supply and manufacture.

Table 2. Requirement of line density

| Line density in Fused Silica | BWW-01 | GWW-01 | RWW-01 |
| --- | --- | --- | --- |
| Conventional design | 6,210 l/mm | 5,280 l/mm | 2,940 l/mm |
| Current design | 5,800 l/mm | 4,920 l/mm | 2,725 l/mm |

Table 3. Comparison of capability

| Design phase | CoDP | LPDP |
| --- | --- | --- |
| Waveband | 0.36-0.90um | 0.36-0.90um |
| Resolution | 40,000 @0.36-0.60um<br>20,000 @0.60-0.90um | 40,000 @0.36-0.50um<br>20,000 @0.50-0.90um |
| Fiber diameter | 0.8" | 0.75" |
| Number of fibers | 578 | 542 |
| Number of spectrographs | 2 | 2 |
| Collimated aperture | Φ300mm | Φ300mm |
| Collimator | F/3.5 on-axis | F/2.05 off-axis |
| Disperser | Grating | Grism+Prism |
| * Opening angle of optical path | 44 degree in BWW/GWW;<br>76 degreein RWW | 67.5 degree in all of three working windows |
| Camera | F/1.6 transmission | F/1.55 transmission |
| Detector | 6Kx6K @15um | 9Kx9K @10um (TBD) |
| Image size | 4.6 pixels | 6 pixels |

* Opening angle of optical path: the angle between incident and output optical axis on the disperser. Its size affects the available space for the mechanical structure of disperser and camera.

# 3. DISPERSER DESIGN

The above section mentions a special disperser design to reduce line density and increase the grating area. In addition, it has two other advantageous functionalities, respectively quickly change the working windows in each spectral channel, and widen the opening angle (up to 67.5 degree) to provide more space for mechanical structure. When the opening angles are identical among Blue, Green and Red channels, it enables matching the difference of spectrum layout at both of resolutions, R=20K and 40K.

The disperser comprises of a special grism and a prism in each spectral channel, see Figure 4. The prisms have the contrary rotational angles between the resolution modes of R=20K and 40K, in order to compensate the difference of opening angles. The special grism is a 'sandwich-like' combination with 2 non-right-angle prisms and 1 grating (surface). The non-right-angle prism takes advantage of the vertex angle to reduce incident angle on air-glass interface, amplify the collimated aperture in dispersed dimension (Y) by refraction effect on entrance surface, and the non-right-angle prism can lighten its self-weight by more than 20%. With assistance to this design, it is not only capable of keeping the fixed location of camera among different working windows, but also has the potential to change the resolution modes by changing the disperser and adjusting the camera focus by a small amount.

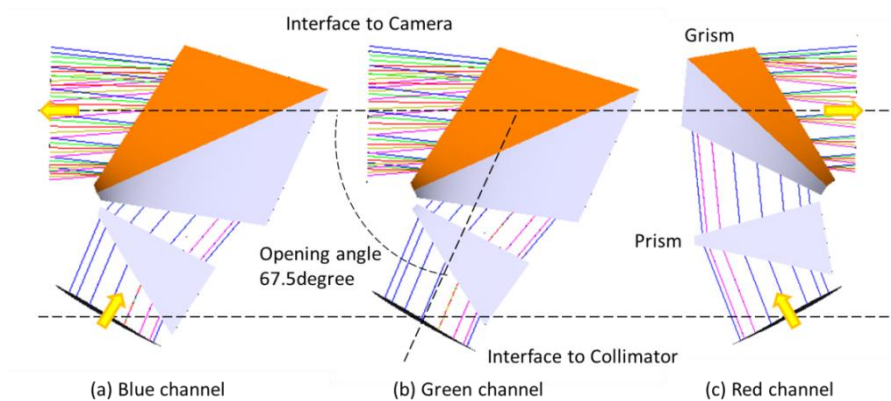

Figure 4. Disperser design

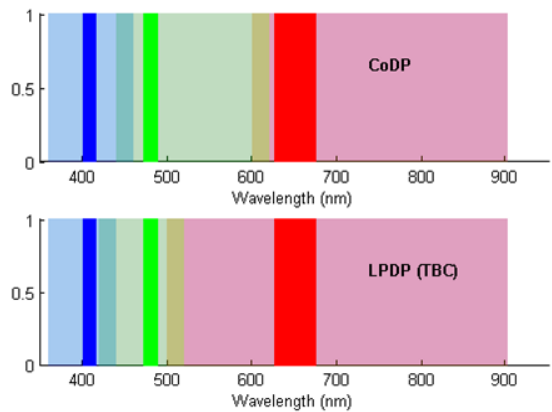

Figure 5. Configuration of spectral channels: (Upper) Configuration at the CoDP; (Lower) Configuration at the LPDP

Blue, green and red narrow bars represent three required working windows; Light blue, light green and light red backgrounds represent the wavelength ranges in three spectral channels.

According to the current requirements, the configuration of spectral channels shown in Figure 5 illustrates the change between the CoDR and LPDP. The red channel bears wider wavelength range than the others in order to overlap the resolution modes in green channel. This disperser design possibly allows switching resolution modes in green channel between R=40,000 and 20,000 as in the CoDP configuration, but it need further checking if overlapping working windows possibly happened in the green channel.

Back to the core question about disperser efficiency, it consists of transmission on AR coatings and diffraction on grating surface. The former item is not much affected by polarization and spatial field angle because all the incident angles at air-glass interface are smaller than 40 degree in the current design. The latter item is the critical problem to figure out with the assistance from the vendors. During our investigation, we paid the most attention on two kinds of grating technologies, VPH and surface-relief gratings. Although VPH grating technology is widely applied in the astronomical spectrographs, only a few vendors can potentially fabricate the required grism due to the ultra-high line density (5,500 – 6,400 l/mm) and oversize grating area. As a result of the current collaboration with some vendors, the theoretical design with line density of 5,500 l/mm for the BWW-01 possibly gets high peak efficiency of 90% at the central wavelength and the minimum efficiency of 60% at the edge wavelength of working window. When line density increases to 6,100 l/mm, the envelope of diffraction efficiency becomes gentler but its peak efficiency reduces to 70%, and minimum efficiency is inevitably lower than 50% at the edge wavelength. These theoretical designs are based on some current manufacturing condition, including the material of grating layer (dichromated gelatin), preparation of holographic grating pattern and so on. It's indicated that the current design with line density of 5800 l/mm in the working window, BWW-01, is capable of getting higher peak efficiency than 80% at the central wavelength and higher minimum efficiency than 50% at the edge wavelength. Meanwhile, this line density selection also concerns minimizing the grating area in the defined range of total line number (N). The corresponding effective grating surface requires a clear aperture of up to 300mm x 580mm, and the physical size of its substrate is necessarily up to 400mm x 700mm, see Figure 4. It needs at least two exposures to cover the adequate area of holographic pattern with reference to the AAT-HERMES (200mm x 500mm) [14]. The other vendors also provide some alternate design concepts, respectively multi-VPH gratings and improved surface-relief grism. The design with improved surface relief technology theoretically gives fantastic diffraction performance, obtaining an average efficiency of high than 90% over the working window. But it is actually in need of more technical confirmation than the VPH grating technology (for details see the reference [13]). The exact grating technology adopted by the MSE HR depends on the further study to confirm the relevant feasibility in manufacture and cost in the ongoing design phase.

## 4. OPTICAL DESIGN

Based on the disperser design, alternate optical systems are investigated to evaluate their synthetic performance in optics and sciences. They are respectively based on off-axis collimator and on-axis collimator. Figure 6 shows the proposed optical design with an off-axis collimator for the MSE HR.

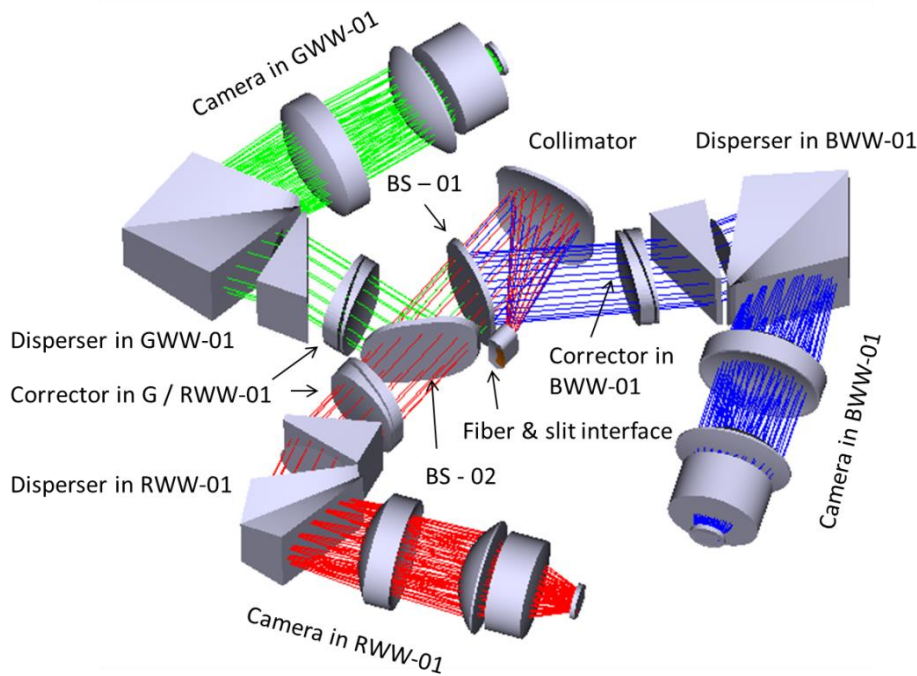

Figure 6. A proposed optical design with an off-axis collimator

(1) FIBER LAYOUT AT SLIT

For off-axis collimator, its most technical advantage is the absence of central obscuration and more freedom for fiber assembly placement, however, it makes fiber assembly geometry more complicated in order to get straight spectrum layout at the detector, see Figure 7. Looked in front view of slit, fiber layout is normally like a 'smile' in order to compensate for the spectrum deviation along the dispersed dimension (Y) on the detector, see Figure 8 (b). The 'smile-like' fiber layout has a radius of 254mm, and its length is 120mm along the spatial dimension (X). It accommodates 542 fibers with central spacing of 220um, and each fiber is 80um in diameter at F/1.926. To fully correct for the field curvature of collimator, hundreds of fibers have to be evenly distributed on a spherical slit surface, see Figure 8 (a). Its spherical radius is about 549mm. Due to off-axis optics, an incident angle of 25.4 degree is added on fiber assembly to match the optical path of off-axis collimator, and it also holds a small gap of 5mm away from the field lens, see Figure 8 (c). Although this kind of fiber assembly has been successfully applied in some MOSs, the fiber assembly works with a F/2.05 fast collimator brings greater challenge than any of them. Fiber assembly for an on-axis collimator does not have requirement of incident angle, and allow aligning all of fiber by touching the last surface of field lens. But its mechanical structure is tightly limited by the central obscuration. So the complexity of fiber assembly is a factor to determine the optimal design for collimator.

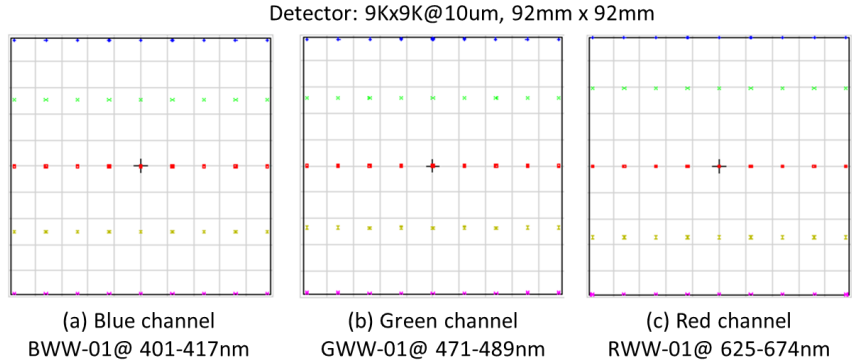

(a) Blue channel  
BWW-01@ 401-417nm

(b) Green channel  
GWW-01@ 471-489nm

(c) Red channel  
RWW-01@ 625-674nm

Figure 7. Spectrum layout at the detector

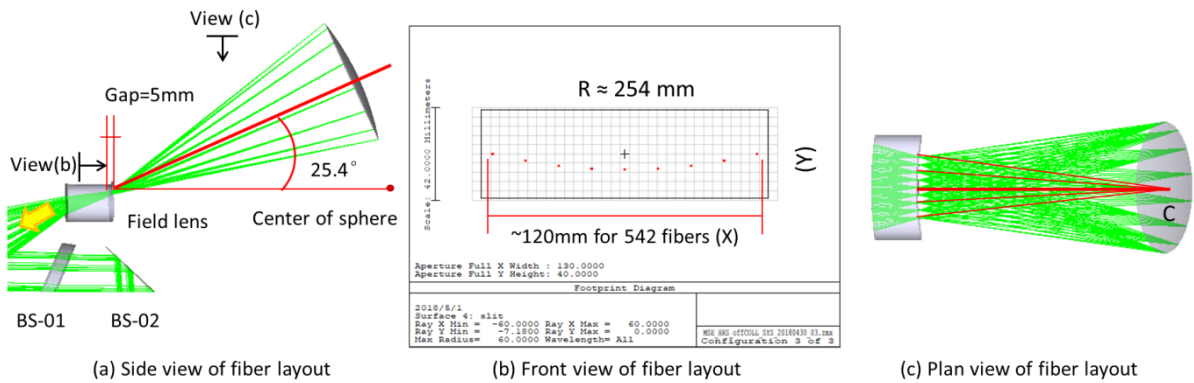

(a) Side view of fiber layout

(b) Front view of fiber layout

(c) Plan view of fiber layout

Figure 8. Fiber layout at slit for the off-axis collimator

(2) COLLIMATOR

The F/2 fast collimator has two options, each holds its own advantage in optical performance and manufacture. The on-axis design adopts the Schmidt system to optimize image quality and throughput by using independent plane corrector in each spectral channel. It holds axial symmetric convenience for optical alignment and lower requirement of asphericity on its corrector than another's. The off-axis design adopts the Houghton system to get similar optical performance by using a pair of aspherical correctors instead of Schmidt corrector, see Figure 9. It removes the spatial limitation for the fiber assembly and avoids the central obscuration. Two dichroic splitters (BS-01, BS-02) locate between the collimating mirror and correctors. Both designs enable good image quality evaluated by a box scale of 80um x 80um, with equivalent to 0.75arcsec x 0.75arcsec on sky, see Figure 10. On the aspect of manufacture, the off-axis design requires one more aspherical corrector, and asphericity of the correctors is also stronger than the on-axis design. As regards throughput, the off-axis design gets 6.5% higher than the on-axis design for the collimator losses shown in Figure 3 and central obscuration of 10%. On the aspect of mechanical accessibility, the on-axis design provides only limited space of 150mm x 15mm x 200mm (X/Y/Z) for the fiber assembly, and the off-axis design enable to give at least 500mm space along the main optical axis (Z). On the aspect of exit pupil, the off-axis system leads an anamorphic pupil with compression ratio of 10% in the dispersed dimension (Y), see Figure 11 (a), which is similar with 4MOST high resolution spectrograph [12]. The disperser design can be used to get a corrected circular pupil, same as the on-axis design, see Figure 11 (b). So this factor doesn't affect the selection of optical design. The proposed design will be further refined by optimize its asphericity and the relevant fabrication processes.

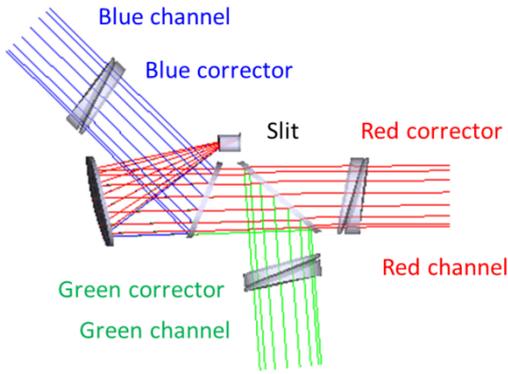

Figure 9. F/2.05 off-axis collimator model

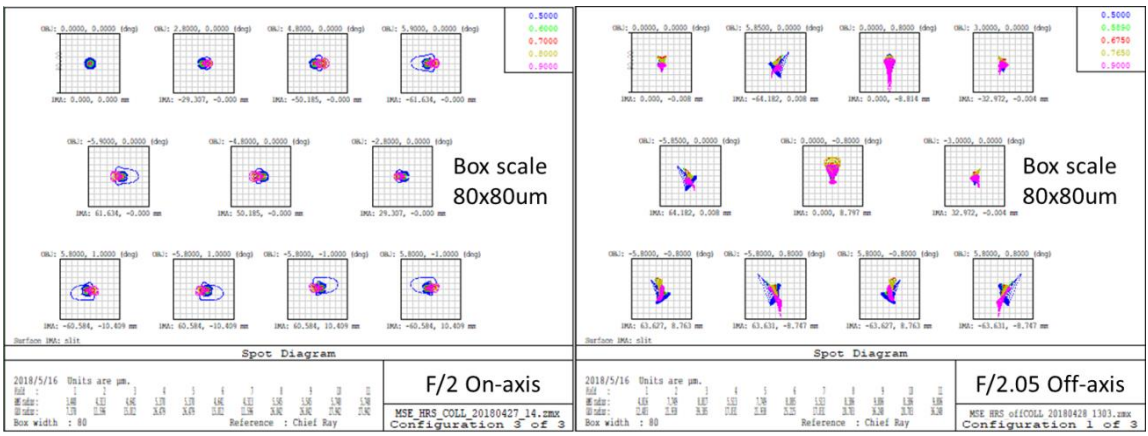

Figure 10. Comparison of spot diagrams in red channel

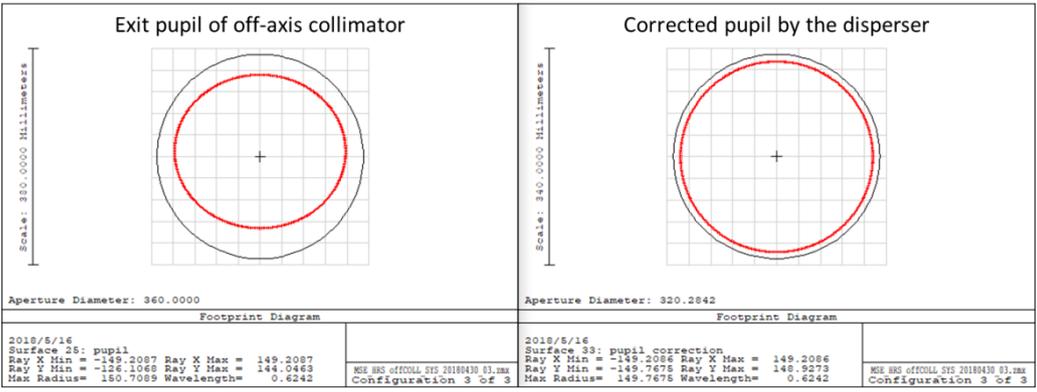

Figure 11. Correction of anamorphic pupil

(3) CAMERA

Each spectral channel feeds a transmission camera, which is composed of a doublet, two singlets and a powered vacuum window. Therein, 3 aspherical surfaces are placed on the first surfaces of 3 lenses marked by asterisk in Figure 12. Among the three cameras, the maximum clear aperture is 500mm in diameter. This greatly limits the available types of transmission glass. The glass types used by the current design consist of S-FSL5Y, S-BSL7, PBM2Y, BSM51Y and Fused Silica. The total physical lengths of cameras are around 900mm (Z). The central spacing between the third lens and the vacuum window is at least wider than 60mm (Z) to accommodate the mounting interface of detector. The central spacing between vacuum window and CCD chip is 7mm (Z). When changes the dispersers to observe in different working windows, the camera in each channel can be quickly re-aligned by adjusting detector's focus and tilt angle about X by ±1.25mm and ±0.05 degree, respectively.

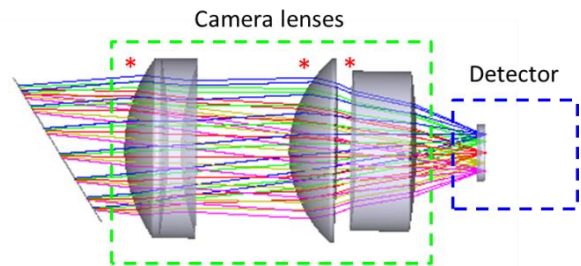

Figure 12. Camera model for the red channel

Integrated with the collimator and dispersers, the complete optical system gets good image quality over the full field of view and the full working windows, see Figure 13. Each box scale is 62um x 62um, with equivalent to 0.75arcsec x 0.75arcsec on sky. The RMS value of spot radius is smaller than 1/4 of geometric image radius, and the GEO value of spot radius is smaller than 1/2 of geometric image radius in all of three working windows. The image quality for any working window in a single spectral channel is approximately identical as to guarantee the similar scientific output. Figure 14 shows the throughput of optical system without disperser. It illustrates that the current design possibly is 7% higher than the design with on-axis collimator, and 15% higher than the conceptual design by comparing with dash curves in Figure 1.

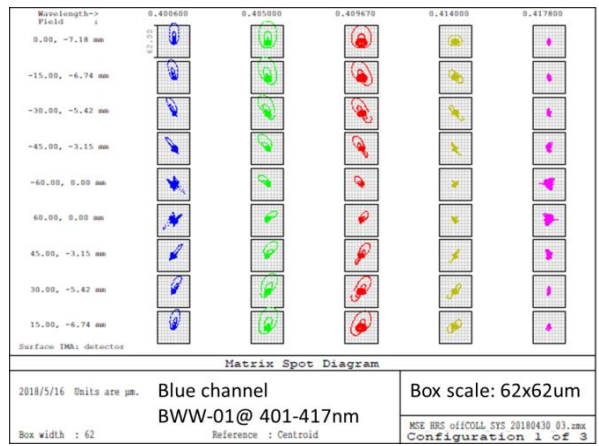

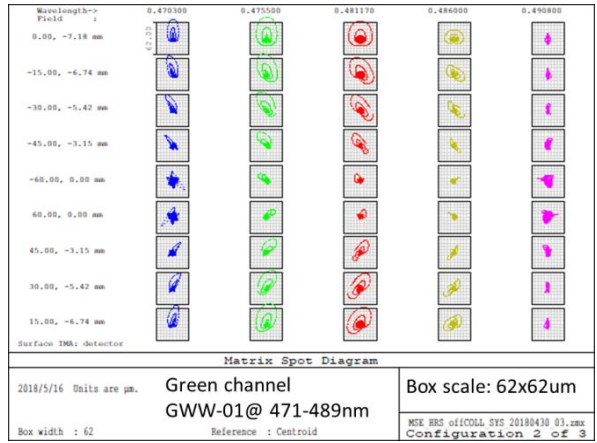

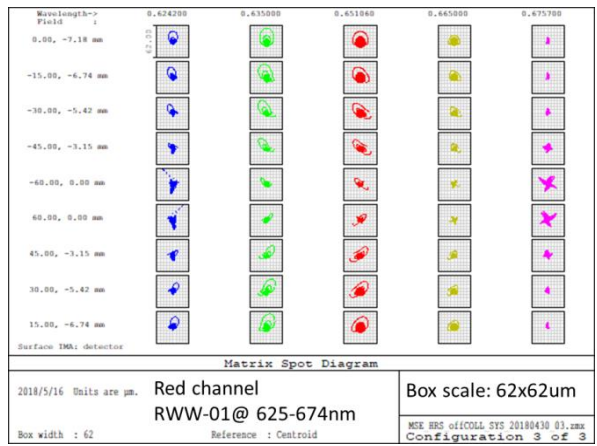

Figure 13. Spot diagrams in three working windows

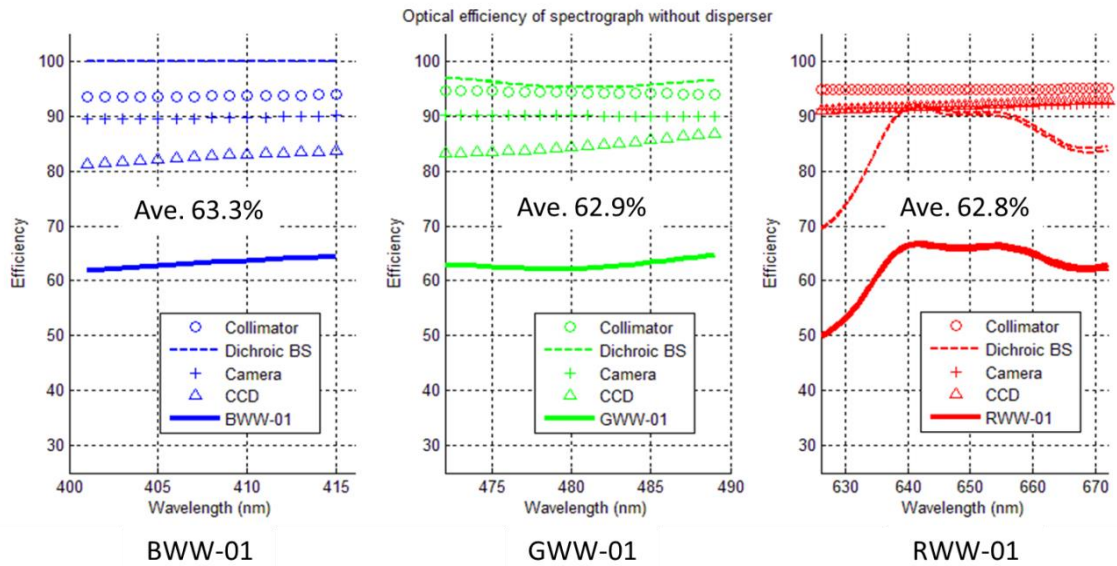

Figure 14. Throughput estimation in three working windows

## 5. SUMMARY

LPDP optical design of the MSE HR is described here to highlight the design challenges of the multi-object spectrographs over the astronomical community currently in the world. The MSE large telescope aperture and its high resolution of 40,000 in the wavelength range of 360-500nm demand to new technological innovation in astronomical instrument scale and ultra-high line density mosaic grism / grating. With in-depth study on the relevant technologies in manufacture, increasingly feasible designs are proposed for this spectrograph. The current design stretches the current manufacturing technologies especially for the disperser and aspherical lenses. In the ongoing design phase, we intend to address the some follow-up issues: (1) The science group to evaluate the scientific impact with lower spectral resolution. (2) The project office to confirm the possibility to change the MSE HR location from the current Coude room to the telescope instrument platforms, in order to increase blue throughput by shortening the fiber length between the prime focus and spectrograph. (3) The HR design group and science group to work together to check which configuration of spectral channels, in terms of wavelength windows, is more reasonable in technology and future science cases. (4) The HR design group and project office to enhance the collaboration with the optical vendors to resolve technical problems on disperser and aspherical lenses.

## ACKNOWLEDGEMENTS

This work is supported by the MSE Project Office, Nanjing Institute of Astronomical Optics and Technology (NIAOT), the National Astronomical Observatory of China (NAOC) and the Chinese Academy of Sciences (CAS). It gets funding support from the Chinese Scholarship Council (CSC) (201704910452, 2017.12-2018.06), the Bureau of Personnel of CAS (2013.9-2014.08), the National Natural Science Foundation of China (NSFC) project (U1531133), (11773047), the Natural Science Foundation of Jiangsu province (BK20151065), the special funds for equipment renewal at astronomical observatory and major equipment operation (2015), the knowledge innovation engineering in young talents field, NIAOT (2014.12-2017.12) and the Youth Innovation Promotion Association, CAS. We sincerely give our thanks to Jim Arns at KOSI, Thomas Flügel-Paul at Fraunhofer IOF, Dominic Speer at Wasatch Photonics, Andrea Bianco at INAF, Christophe Gombaud at HORIBA Scientics, Turan Erdogan at Plymouth Grating Laboratory, Inc and so on. All of them gave great technical support and professional advice on disperser. We appreciate that Derrick Salmon, Andrew Sheinis at the CFHT, and Shan B. Mignot at Observatoire de Paris gave much meaningful comment during the design phase, Prof. Xuefei Gong at NIAOT gave the HR design group full support in funding, international cooperation and project management.

## REFERENCES

[1] Doug A. Simons, et al., "Current status and future plans for the Maunakea Spectroscopic Explorer (MSE)," Proc. SPIE. 9145, 914515 (2014).

[2] Kei Szeto, et al., "Maunakea Spectroscopic Explorer Design Development from Feasibility Concept to Baseline Design," Proc. SPIE 9906, 99062J (2016).

[3] Alan W. McConnachie, et al., "Science-based requirements and operations development for the Maunakea Spectroscopic Explorer," Proc. SPIE 9906, 99063 (2016).

[4] Kei Szeto, et al., "Maunakea spectroscopic explorer emerging from conceptual design," Proc. SPIE 10700, 54 (2018).

[5] Nicolas Flagey, et al., "The Maunakea Spectroscopic Explorer: throughput optimization," Proc. SPIE 9908, 99089C (2016).

[6] Kai Zhang, et al., "Mauna Kea Spectrographic Explorer (MSE): a conceptual design for multi-object high resolution spectrograph," Proc. SPIE 9908, 99081P (2016).

[7] Samuel C. Barden, et al., "HERMES: revisions in the design for a high-resolution multi-element," Proc. SPIE 7735, 773509 (2010).

[8] Kaiyuan Zhang, et al., "High Numerical Aperture Multimode Fibers for Prime Focus Use," Proc. SPIE 9912, 99125J (2016).

[9] Sébastien Vives, et al., "A spectrograph instrument concept for the Prime Focus Spectrograph (PFS) on Subaru Telescope," Proc. SPIE. 8446, 84464P (2012).

[10] E. Oliva, et al., "Updated optical design and trade-off study for MOONS, the Multi-Object Optical and Near Infrared spectrometer for the VLT," Proc. SPIE 9147, 91472C (2014).

[11] Gavin Dalton, et al., "Project overview and update on WEAVE: the next generation wide-field spectroscopy facility for the William Herschel Telescope," Proc. SPIE 9147, 91470L (2014).

[12] W. Seifert, et al., "4MOST: The High-Resolution Spectrograph," Proc. SPIE 9908, 990890 (2016).

[13] Will Saunders, et al., "Higher dispersion and efficiency Bragg gratings for optical spectroscopy," Proc. SPIE 10706, 187 (2018).

[14] J. A. C. Heijmans, et al., "Design and development of the high-resolution spectrograph HERMES and the unique volume phase holographic gratings," Proc. SPIE 8167, 81671A (2011).